\documentclass{article}
\usepackage{spconf,amsmath,graphicx}

\title{Voice Aging with Audio-Visual Style Transfer} 
\name{Justin Wilson*$^1$, Sunyeong Park*$^3$,  Seunghye J. Wilson$^3$, Ming C. Lin$^{2,1}$ 
}
\address{
{$^{1}$Department of Computer Science, UNC Chapel Hill, United States} \\
{$^{2}$Department of Computer Science, UMD College Park, United States} \\
{$^{3}$Independent Researcher}
}
\begin{document}
%
\maketitle
\begin{abstract}
Face aging techniques have used generative adversarial networks (GANs) and style transfer learning to transform one's appearance to look younger/older. Identity is maintained by conditioning these generative networks on a learned vector representation of the source content. In this work, we apply a similar approach to age a speaker's voice, referred to as `voice aging'. We first analyze the classification of a speaker's age by training a convolutional neural network (CNN) on the speaker's voice and face data from Common Voice and VoxCeleb datasets. We generate aged voices from style transfer to transform an input spectrogram to various ages and demonstrate our method on a mobile app.
\end{abstract}

\section{Introduction}
Large-scale audio datasets like Common Voice~\cite{CommonVoice} and video datasets such as VoxCeleb~\cite{Nagrani17,Chung18b} have enabled deep learning neural networks to predict speakers' ages based on their voice and/or face to varying degrees of granularity and accuracy. Generative adversarial networks (GANs) have been developed to generate younger or older faces conditioned on a source image~\cite{DBLP:journals/corr/AntipovBD17}. 
These generative models are referred to as face aging. The FaceApp is a mobile application that recently went viral for its face aging and transformations (e.g. emotion, gender, etc.). Since launching in 2017, FaceApp was downloaded by more than an estimated 86 million users worldwide, according to Business Insider in 2019.

In addition to image data, audio of a speaker has also been used alone and in conjunction with visual information for biometric identification, speaker recognition, etc.~\cite{Chaudhari2015AutomaticSA}. For example, segment duration and sound pressure level (SPL) range were shown to be acoustic correlates of speaker age~\cite{Muller2003}. Cross-modal tasks have also be demonstrated. For example, recognizing face based on voice and vice versa~\cite{Nagrani18a,Nagrani18c}. Voice impersonation~\cite{DBLP:journals/corr/abs-1802-06840} or voice conversion~\cite{fang2018highquality} using generative adversarial networks has also been studied. 

\begin{figure}[thpb]
  \centering
  \includegraphics[width=0.5\textwidth]{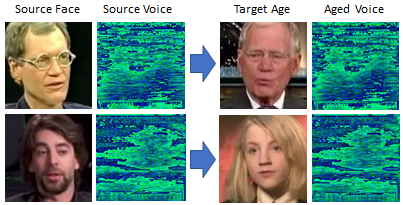}
  \caption{We use CycleGAN transfer learning to age a voice from young to old (top row) or old to young (bottom row). A desired target age can be given or predicted based on the target speaker's face and voice using our Voice Aging Neural Network (VANN). We evaluate our method using VoxCeleb1 and develop a mobile app for data collection and playback.}
  \label{fig:fig1}
\end{figure}

Our voice aging method generates audio conditioned on a target age by using style transfer of a CNN based on audio and audio-visual data. Our main contributions include:

\begin{enumerate}
    \item Annotated VoxCeleb1 data with age based on YouTube video information and celebrity birth date;
    \item Trained audio and audio-visual age classification VANN neural networks based on speaker voice and face;
    \item Translated voice from young to old and old to young via CycleGAN style transfer.
\end{enumerate}

\section{Background and Related Work}
While perhaps less noticeable than facial features, our voices change as we age, especially noticeable from youth as our hormone levels change and larynx grows~\cite{Hollien1960,Wells13}. There is also evidence of voice changing in adulthood as a consequence of losing muscle mass, thinning mucous membranes, lessening lung capacity, and changing voice box~\cite{DukeHealthWeb}. These differences in voice aging can be caused by vocal cord atrophy or bowing known as presbyphonia or presbylaryngeus~\cite{DukeHealthWeb,HarvardHealthWeb}. Aging voice has also been shown to impact the performance of speech recognition systems, highlighting a need for software to account for this change~\cite{Vipperla2010,Aman2013AnalyzingTP}.

\textbf{Biometric identification}: recognition systems have been developed to identify individuals based on such audio and visual traits. Identification systems commonly use biometrics such as face recognition and voice identification and must be aware of their dynamic nature as they change over time. To predict demographics of the speaker such as gender, age, and ethnicity, features from audio~\cite{Schtz2007ASO}, visual~\cite{Bhatia13}, and audio-visual~\cite{Nagrani18a,Gad15} information have been studied. 

\textbf{Style transfer learning}: takes a source and target image and generates an output image resembling the source but styled as the target image~\cite{DBLP:journals/corr/GatysEB15a}. One approach for image style transfer has been to use Convolutional Neural Networks (CNNs)~\cite{Gatys2016}. Style transfer is similar to GANs as they both generate outputs to minimize loss. The difference lies in the objective loss -- for GANs, it is realism; for style transfer it is based on content, style, and total variation loss.

\textbf{Face transformation}: 
face aging research has been achieved by using the encoded vector representation of the subject's face as an input into the generator, along with the desired age. 
The objective is to obtain a face transformation while conditioning on the source identity~\cite{DBLP:journals/corr/AntipovBD17}. Conditional adversarial autoencoders have been used~\cite{zhang2017age}, as well as realistic neural talking head models that use few shot adversarial learning and encoded facial landmarks~\cite{DBLP:journals/corr/abs-1905-08233}.

\textbf{Audio transformation}: research has been conducted to evaluate the impact of voice aging on Automatic Speech Recognition~\cite{Vipperla2010,Aman2013AnalyzingTP} and its consideration for medical diagnosis~\cite{DukeHealthWeb,HarvardHealthWeb}. Statistics of "auditory textures" have also been used to classify environmental acoustic events (e.g. water, rain, fire, etc.) assuming stationarity. Synthetic signals can emulate these sounds by matching the statistics of the textures~\cite{McDermott09,McDermott11,McDermott13}. Audio based style transfers have also been performed on spectrograms~\cite{DBLP:journals/corr/abs-1710-11385,DBLP:journals/corr/abs-1801-01589} as well as GANs~\cite{DBLP:journals/corr/abs-1802-04208,pasini2019melganvc}.  

\section{Technical Approach}
\label{TechnicalApproach}

We use a multimodal CNN (Fig.~\ref{fig:network_figure}) and CycleGAN (Fig.~\ref{fig:gan_figure}) for voice aging with age prediction and style transfer.

\subsection{Datasets}
\label{Datasets}
\begin{figure}[thpb]
  \centering
  \includegraphics[width=0.4\textwidth]{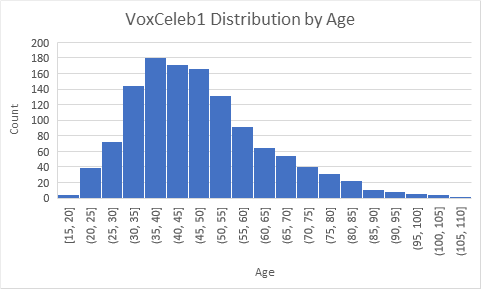}
  \caption{VoxCeleb1 dataset distribution of speakers by age. 71\% of speakers are between 30-60 years of age. The male to female ratio is 55\% to 45\% respectively. 81\% of VoxCeleb1 celebrities are from the United States (64\%) and United Kingdom (17\%). Please note that this figure is count by speaker, although distribution for video count is similar because we limit number of videos per person.}
  \label{fig:voxCeleb1AgeDist}
\end{figure}

The Common Voice~\cite{CommonVoice} dataset was used to test age classification using voice alone as the dataset is audio only. It is comprised of a short sentence being read, only a few seconds in length. Each speaker has associated age, gender, and nationality metadata. VoxCeleb1~\cite{Nagrani17} was used to test audio and audio-visual age classification and voice aging. Unlike Common Voice, VoxCeleb1 does not have age metadata associated with the video data. We annotate age for each video based on celebrity birthday and YouTube video recorded date obtained from the title, description, or published date. Below is a distribution of the VoxCeleb1 data by age (Fig.~\ref{fig:voxCeleb1AgeDist}).

Note that some of the VoxCeleb1 videos are no longer available, because the YouTube account associated with the video has been terminated, the video contains content that has been blocked on copyright grounds, or the video is private. This accounted for approximately 8\% of possible videos.

\subsection{Audio-visual age classification}
\label{AudioVisualAgeClassification}

\begin{figure*}
  \centering
  \includegraphics[width=0.99\textwidth]{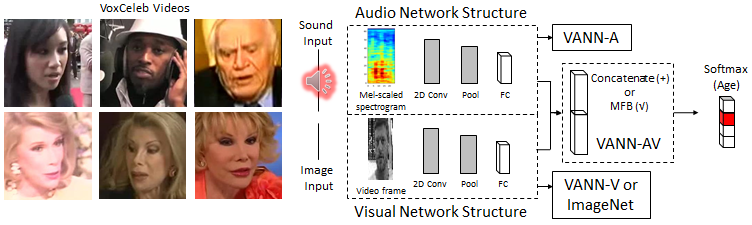}
  \caption{Ground truth age labels are prepared from VoxCeleb videos based on celebrity birthday and video recorded date. Cropped images of faces~\cite{Nagrani18a} and intervals of audio converted into mel-scaled spectrogram are input into our multimodal CNN, referred to as Voice Aging Neural Network (VANN). Classification is used for A, B and $\leq$25, 26-50, 51-75, and $>$75 years old.}
  \label{fig:network_figure}
\end{figure*}

A desired target age can be given or predicted based on the target speaker's face and voice using our Voice Aging Neural Network (VANN). Our audio-based variant VANN-A, shown in Fig.~\ref{fig:network_figure}, consists of a single convolutional layer followed by two dense layers with feature normalization. It is trained on mel-scaled spectrograms for audio intervals of 0.24 seconds and  performs optimally on our classification tasks (Table ~\ref{tab:model-eval}).

\begin{figure}[thpb]
  \centering
  \includegraphics[width=0.5\textwidth]{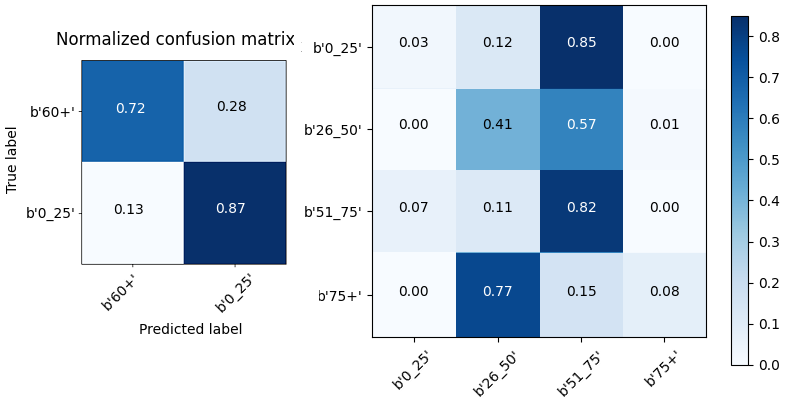}
  \caption{Confusion matrices of age classification based on face and voice data. 
  (Left) Young, A and old, B form the basis for CycleGAN voice aging. (Right) Incorrect predictions of age classes 26-75 
  may be due to insufficient or imbalanced data.}
  \label{fig:confMarix}
\end{figure}

\textbf{Results}: while initial results suggest a benefit from audio-visual age prediction based on face and voice and form a basis for voice aging style transfer, there remain areas for improvement. Fixing imbalanced data as shown in Fig.~\ref{fig:voxCeleb1AgeDist} may further improve our neural network's ability to distinguish between granular age classes (Fig.~\ref{fig:confMarix}). Furthermore, feature engineering to split training and test sets by age and/or ethnicity as well as an ablative study for model selection with deeper and sequential layers are other areas for future work.

\subsection{Voice aging style transfer}
\label{VoiceAgingStyleTransfer}

\textbf{Pre-processing}: our research goal is to apply the concept of GAN style transfer to human voices to make them sound older or younger. Voice data can be composed of three dimensions of frequency, amplitude, and time, and can be well represented through spectrograms. During pre-processing, audio intervals from VoxCeleb videos are converted into spectrogram images to serve as input data into a CycleGAN trained from existing voice data. VoxCeleb's voice data is split into several wav files of 0.24 seconds, about the length of a syllable, with each interval converted into a mel-spectrogram.

To create images that are the same height and width to use as input into a CycleGAN, the frequency y-axis length and time x-axis are defined as 128 by 128 pixels. The amplitudes are converted into RGB-dimensional color values. Each RGB value generated in Equation~\ref{eqn:eqnSpec} is changed to a natural number less than 256, discarding all digits after the decimal point.

\begin{equation}
\label{eqn:eqnSpec}
    S(x, y) = log(amplitude(x, y)) * scale
\end{equation}

\noindent where Green(x,y)$=(S(x, y)\text{ mod } 256^2) / 256$, Red(x,y)$=S(x, y)/256^2$, and Blue(x,y)$=(S(x, y)\text{ mod }256)$. In order to make the change in amplitude as homogeneous as possible, log is used as the amplitude. By taking the inverse function of the generated image data again, it can be easily converted into an original audio file with a slight loss due to conversions and reconstruction.

\begin{table}
\centering
  \begin{tabular}{l|c|ccc}
  \multicolumn{5}{c}{Age Estimation Accuracy (Acc) by Method and Input} \\
  \hline
    Class by age range every & & 10 yr & 25 yr & A, B \\
    
    Method        & In               & Acc & Acc & Acc  \\ 
    \hline
    K-Nearest Neighbors
    & A & 20.2\% & 36.8\% & 62.6\%  \\
    Linear SVM
    & A & 14.1\% & 23.3\% & 58.1\%  \\
    SoundNet8~\cite{aytar2016soundnet} & A & 19.3\% & 45.9\% & 67.1\% \\
    
    \textbf{VANN-A} (Ours) & A & 24.7\% & 45.9\% & 71.0\% \\
    \hline
    ImageNet~\cite{ImageNet} & V & 24.7\% & 50.0\% & 57.4\% \\
    \textbf{VANN-V} (Ours) & V & 22.3\% & 50.2\% & 77.4\% \\
    \hline
    \textbf{VANN-AV Cat} (Ours) & AV & 24.7\% & 46.0\% & 70.4\% \\
    \textbf{VANN-AV MFB} (Ours) & AV & \textbf{26.1\%} & \textbf{52.7\%} & \textbf{80.4\%} \\
    \hline
  \end{tabular}
  \caption{Datasets were trained for 15, 30, and 45 videos per VoxCeleb totaling 18,157 (82\%), 36,312 (90\%), and 54,460 (93\%) pairs of face and voice data. The hold out test set consisted of 4,000 pairs. Training was done on a Titan X GPU for 40 epochs and batch size of 32. $A \leq 25$ and $B > 60$.  Our methods (bold-faced) perform at least equally well or better than other alternatives.}
  \label{tab:model-eval}
\end{table}




\begin{figure*}[h]
  \centering
  \includegraphics[width=0.84\textwidth]{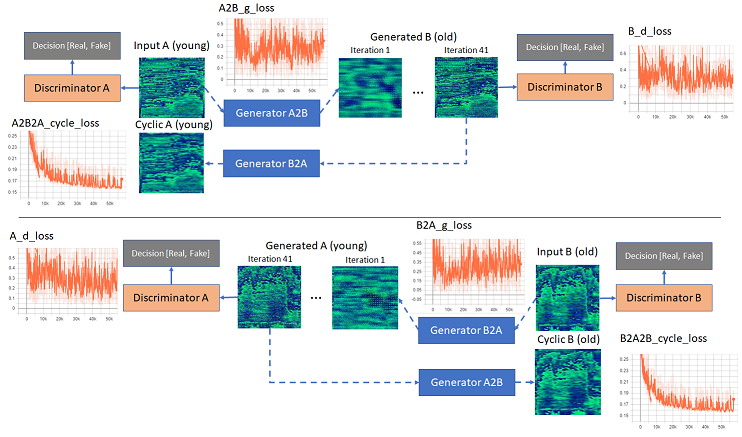}
  \caption{We use a CycleGAN to perform voice-to-voice translations from young A (18-25) to old B (above 60) and vice versa. It is trained for 50 epochs on 0.24 second audio intervals from VoxCeleb1 video dataset. CycleGAN maintains source content by relying on pixel-wise loss while transforming between similar domains, making it a suitable approach for our voice aging task.}
  \label{fig:gan_figure}
\end{figure*}

\textbf{Training}: the objective is to pronounce syllables of all ages for everyone included in the dataset that has implemented style transfer over a common GAN network. However, it is difficult to obtain class data for all classes, so we apply the concept of CycleGAN. The voice data is divided into three groups: young, middle, and old, and style transmission converts the voice of the young to the voice of the old. All models were implemented with Tensorflow and Keras. 

\begin{enumerate}
    \item Generator $G$ ages the spectrogram image of a young person ($A \leq 25\text{ yrs}$) into a spectrogram image of the old age ($B > 60\text{ yrs}$)
    \item Generator $F$ ages the spectrogram image of old age ($B > 60$) into the spectrogram image of young man ($A \leq 25\text{ yrs}$)
    \item Discriminator $D_A$ distinguishes the authenticity of A
    \item Discriminator $D_B$ distinguishes the authenticity of B
\end{enumerate}

\begin{figure}[htb]
  \centering
  \includegraphics[width=0.5\textwidth]{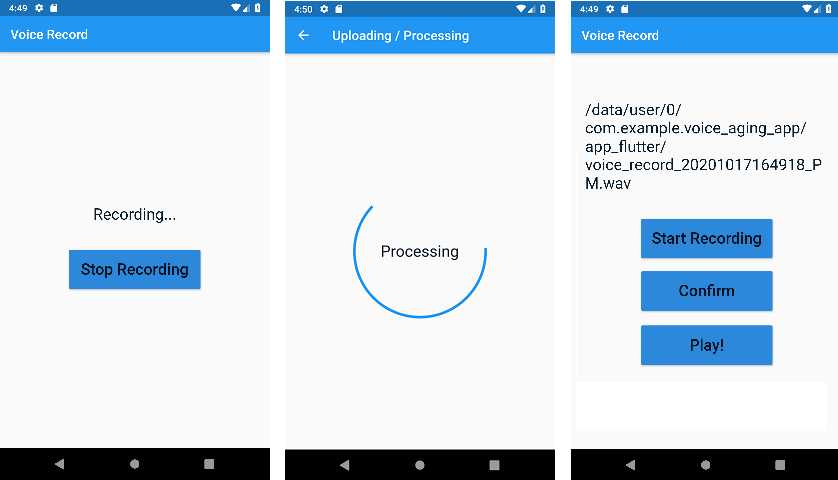}
  \caption{Our voice aging app demonstrates the ability of a user to record their voice and playback their younger and older self. Recorded audio is sent to a server for spectrogram processing and age translation.}
  \label{fig:mobileApp}
\end{figure}

Our CycleGAN model is implemented by letting $G$ generate B from A and then again $F$ to generate the data as A. Fig.~\ref{fig:gan_figure} displays generated images at iteration 1 and 41 as well as discriminator, generator, and cycle losses over 50 epochs of training. By using this trained model, the voice recorded in a mobile app (Fig.~\ref{fig:mobileApp}) is sent to a server where this model is uploaded, processed, and aged to make the voice older or younger. The app can also be used in the future for additional data collection and training.

\section{Conclusion and Future Work}
\label{Conclusion}

Like face aging research, we apply a generative spectrogram-to-spectrogram translation to age a speaker's voice, we refer to as voice aging. Given a target face and voice, we predict the desired age based on our trained voice aging convolutional neural network, called VANN. With audio-visual prediction of target age, we use this to form a basis for voice aging where we generate aged voices using style transfer to transform an input spectrogram to various ages. Our models are trained and evaluated on speaker data from Common Voice and VoxCeleb datasets. We demonstrate the application of our method on a mobile app sends and receives recorded audio to a server for spectrogram processing, style transfer, and playback. 

\textbf{Future work}: while initial results suggest promising results from our voice aging network, there are areas for further research. For instance, fixing insufficient or imbalanced datasets by using a number of additional audio-visual datasets such as Lip Reading Datasets (LRW,LRS2,LRS3), VoxCeleb2 containing about 5,000 more speakers than VoxCeleb1, MIT Mobile Device Speaker Verification Corpus~\cite{Woo06}, and speaker-independent audio-visual dataset for speech separation~\cite{DBLP:journals/corr/abs-1804-03619}. Also, feature engineering to split these datasets by age and/or ethnicity as well as evaluating against other deep learning models with more layers and recurrent structures are other areas for future work. Further investigation into age prediction accuracy and generation should be explored to 
enable more granular age prediction and aging.

\newpage

\begin{small}
\bibliographystyle{IEEEbib}
\bibliography{strings,refs}

\begin{thebibliography}{10}

\bibitem{CommonVoice}
Kamp~T. Henretty, M. and K.~Davis,
\newblock ``Common voice,'' https://voice.mozilla.org/en.

\bibitem{Nagrani17}
A.~Nagrani, J.~S. Chung, and A.~Zisserman,
\newblock ``Voxceleb: a large-scale speaker identification dataset,''
\newblock in {\em INTERSPEECH}, 2017.

\bibitem{Chung18b}
J.~S. Chung, A.~Nagrani, and A.~Zisserman,
\newblock ``Voxceleb2: Deep speaker recognition,''
\newblock in {\em INTERSPEECH}, 2018.

\bibitem{DBLP:journals/corr/AntipovBD17}
Grigory Antipov, Moez Baccouche, and Jean{-}Luc Dugelay,
\newblock ``Face aging with conditional generative adversarial networks,''
\newblock {\em CoRR}, vol. abs/1702.01983, 2017.

\bibitem{Chaudhari2015AutomaticSA}
Shivaji~J Chaudhari and Ramesh~Mahadev Kagalkar,
\newblock ``Automatic speaker age estimation and gender dependent emotion
  recognition,''
\newblock 2015.

\bibitem{Muller2003}
Christian Müller, Frank Wittig, and Jörg Baus,
\newblock ``Exploiting speech for recognizing elderly users to respond to their
  special needs.,''
\newblock 01 2003.

\bibitem{Nagrani18a}
A.~Nagrani, S.~Albanie, and A.~Zisserman,
\newblock ``Seeing voices and hearing faces: Cross-modal biometric matching,''
\newblock in {\em IEEE Conference on CVPR}, 2018.

\bibitem{Nagrani18c}
A.~Nagrani, S.~Albanie, and A.~Zisserman,
\newblock ``Learnable pins: Cross-modal embeddings for person identity,''
\newblock in {\em European Conference on Computer Vision}, 2018.

\bibitem{DBLP:journals/corr/abs-1802-06840}
Yang Gao, Rita Singh, and Bhiksha Raj,
\newblock ``Voice impersonation using generative adversarial networks,''
\newblock {\em CoRR}, vol. abs/1802.06840, 2018.

\bibitem{fang2018highquality}
Fuming Fang, Junichi Yamagishi, Isao Echizen, and Jaime Lorenzo-Trueba,
\newblock ``High-quality nonparallel voice conversion based on cycle-consistent
  adversarial network,'' 2018.

\bibitem{Hollien1960}
H.~Hollien and Moore~G. P.,
\newblock ``Measurements of the vocal folds during changes in pitch,''
\newblock in {\em Journal of Speech, Language, and Hearing Research}, 1960, pp.
  157--165.

\bibitem{Wells13}
Baguley T. Sergeant~M. Wells, T. and A.~Dunn,
\newblock ``Perceptions of human attractiveness comprising face and voice
  cues,'' 2013.

\bibitem{DukeHealthWeb}
Duke Health,
\newblock ``Aging voice problems,''
  https://www.dukehealth.org/treatments/voice-disorders/aging-voice.

\bibitem{HarvardHealthWeb}
Harvard Health,
\newblock ``Aging voice,''
\newblock {\em Harvard Health Publishing}, vol.
  https://www.health.harvard.edu/staying-healthy/aging-voice, 2013.

\bibitem{Vipperla2010}
R.~Vipperla, S.~Renals, and J.~Frankel,
\newblock ``Ageing voices: The effect of changes in voice parameters on asr
  performance,''
\newblock in {\em Journal on Audio, Speech, and Music Processing}. EURASIP,
  2010.

\bibitem{Aman2013AnalyzingTP}
Fr{\'e}d{\'e}ric Aman, Michel Vacher, Solange Rossato, and François Portet,
\newblock ``Analyzing the performance of automatic speech recognition for
  ageing voice: Does it correlate with dependency level?,''
\newblock in {\em SLPAT}, 2013.

\bibitem{Schtz2007ASO}
Susanne Sch{\"o}tz and Christian~A. M{\"u}ller,
\newblock ``A study of acoustic correlates of speaker age,''
\newblock in {\em Speaker Classification}, 2007.

\bibitem{Bhatia13}
R.~Bhatia,
\newblock ``Biometrics and face recognition techniques,''
\newblock in {\em International Journal of Advanced Research in Computer
  Science and Software Engineering}, 2013, pp. 93--99.

\bibitem{Gad15}
El-Fishawy N. EL-SAYED~A. Gad, R. and M.~Zorkany,
\newblock ``Multi-biometric systems: A state of the art survey and research
  directions,'' 2015.

\bibitem{DBLP:journals/corr/GatysEB15a}
Leon~A. Gatys, Alexander~S. Ecker, and Matthias Bethge,
\newblock ``A neural algorithm of artistic style,''
\newblock {\em CoRR}, vol. abs/1508.06576, 2015.

\bibitem{Gatys2016}
Ecker A. S.-Bethge~M. Gatys, L.~A.,
\newblock ``Image style transfer using convolutional neural networks,''
\newblock in {\em IEEE Conference on Computer Vision and Pattern Recognition
  (CVPR)}, 2016.

\bibitem{zhang2017age}
Zhifei Zhang, Yang Song, and Hairong Qi,
\newblock ``Age progression/regression by conditional adversarial
  autoencoder,''
\newblock in {\em IEEE Conference on CVPR}, 2017.

\bibitem{DBLP:journals/corr/abs-1905-08233}
Egor Zakharov, Aliaksandra Shysheya, Egor Burkov, and Victor~S. Lempitsky,
\newblock ``Few-shot adversarial learning of realistic neural talking head
  models,''
\newblock {\em CoRR}, vol. abs/1905.08233, 2019.

\bibitem{McDermott09}
Oxenham A. Simoncelli~E. McDermott, J.,
\newblock ``Sound texture synthesis via filter statistics,''
\newblock in {\em IEEE Workshop on Applications of Signal Processing to Audio
  and Acoustics}, 2009.

\bibitem{McDermott11}
J.~McDermott and E.~Simoncelli,
\newblock ``Sound texture perception via statistics of the auditory periphery:
  Evidence from sound synthesis,'' 2011.

\bibitem{McDermott13}
Schemitsch~M. McDermott, J. and E.~Simoncelli,
\newblock ``Summary statistics in auditory perception,'' 2013.

\bibitem{DBLP:journals/corr/abs-1710-11385}
Eric Grinstein, Ngoc Q.~K. Duong, Alexey Ozerov, and Patrick P{\'{e}}rez,
\newblock ``Audio style transfer,''
\newblock {\em CoRR}, vol. abs/1710.11385, 2017.

\bibitem{DBLP:journals/corr/abs-1801-01589}
Prateek Verma and Julius O.~Smith III,
\newblock ``Neural style transfer for audio spectograms,''
\newblock {\em CoRR}, vol. abs/1801.01589, 2018.

\bibitem{DBLP:journals/corr/abs-1802-04208}
Chris Donahue, Julian~J. McAuley, and Miller~S. Puckette,
\newblock ``Synthesizing audio with generative adversarial networks,''
\newblock {\em CoRR}, vol. abs/1802.04208, 2018.

\bibitem{pasini2019melganvc}
Marco Pasini,
\newblock ``Melgan-vc: Voice conversion and audio style transfer on arbitrarily
  long samples using spectrograms,'' 2019.

\bibitem{aytar2016soundnet}
Yusuf Aytar, Carl Vondrick, and Antonio Torralba,
\newblock ``Soundnet: Learning sound representations from unlabeled video,''
\newblock in {\em Advances in NIPS}, 2016.

\bibitem{ImageNet}
J.~Deng, W.~Dong, R.~Socher, L.-J. Li, K.~Li, and L.~Fei-Fei,
\newblock ``{ImageNet: A Large-Scale Hierarchical Image Database},''
\newblock in {\em CVPR09}, 2009.

\bibitem{Woo06}
Park~A. Woo, R. and T.~J. Hazen,
\newblock ``The mit mobile device speaker verification corpus: Data collection
  and preliminary experiments,''
\newblock in {\em Proceedings of Odyssey 2006, The Speaker and Language
  Recognition Workshop}, 2006.

\bibitem{DBLP:journals/corr/abs-1804-03619}
Ariel Ephrat, Inbar Mosseri, Oran Lang, Tali Dekel, Kevin Wilson, Avinatan
  Hassidim, William~T. Freeman, and Michael Rubinstein,
\newblock ``Looking to listen at the cocktail party: {A} speaker-independent
  audio-visual model for speech separation,''
\newblock {\em CoRR}, vol. abs/1804.03619, 2018.

\end{thebibliography}
\end{small}

\end{document}